\documentclass[twocolumn,aps,prl,showpacs,superscriptaddress]{revtex4}
\usepackage[dvips]{graphicx}
\usepackage{bm}
\begin{document}
\title{Theory of high-symmetry tetramer single molecule magnets}
\author{Richard A. Klemm}
\email{klemm@phys.ksu.edu} \affiliation{Department of Physics,
Kansas State University, Manhattan, KS 66506 USA}
\author{Dmitri V. Efremov}
\email{efremov@theory.phy.tu-dresden.de} \affiliation{Institut
f{\"u}r Theoretische Physik, Technische Universit{\"a}t Dresden,
01062 Dresden, Germany}
\date{\today}
\begin{abstract}
We present a microscopic theory of single molecule magnets. From
our exact single-ion spin matrix elements for four arbitrary
spins, we study the single-ion anisotropy of equal spins
exhibiting $T_d$, $D_{2d}$, or $C_{4v}$ molecular group symmetry.
Each  group generates site-dependent single-ion anisotropy. For
weak anisotropy, accurate  Hartree expressions
  for the magnetization, specific heat, electron
paramagnetic resonance (EPR) absorption and inelastic neutron
scattering cross-section are given. For $D_{2d}$, azimuthal
single-ion anisotropy leads to the observed Ni$_4$ EPR splittings.
\end{abstract}
 \vskip0pt\vskip0pt
\maketitle

Single molecule magnets (SMM's) have been a topic of great
interest for more than a decade,\cite{general} because of their
potential uses in quantum computing and/or magnetic
storage,\cite{moregeneral} which are possible due to magnetic
quantum tunneling  (MQT) and entangled states.  In fits to a
wealth of data,  the Hamiltonian within an SMM was assumed to be
the  Heisenberg exchange interaction plus weaker global (total, or
giant) spin anisotropy interactions, with a fixed overall global
spin quantum number $s$.\cite{general} MQT and entanglement were
studied in this simple model.

The simplest SMM's  are dimers.\cite{ek,ek2} Surprisingly, two
antiferromagnetic dimers, an Fe$_2$ and a Ni$_2$, appear to have
substantial single-ion anisotropy without any appreciable global
anisotropy.\cite{Shapira,Mennerich,ek2} Although the most common
SMM's have ferromagnetic (FM) intramolecular interactions and
contain $n\ge8$ magnetic ions,\cite{Dalal} a number of
intermediate-sized FM SMM's with $n=4$ and rather simple molecular
structures were recently studied.   Co$_4$ and Cr$_4$ have $s=6$
ground states with spin 3/2 ions  on the corners of
tetrahedrons.\cite{Co4,Cr4}  A number of high symmetry $s=4$
ground state Ni$_4$ structures with spin 1 ions were
reported.\cite{Ni4,Edwards,Boskovic}  Inelastic neutron scattering
(INS) experiments provided strong evidence for single-ion
anisotropy in Co$_4$ and a Ni$_4$.\cite{Ni4,Co4} Fits to electron
paramagnetic resonance (EPR) Ni$_4$ data assuming a fixed $s$ were
also problematic.\cite{Hillprivate}

Yet there is no microscopic model of FM SMM's in which the MQT and
entanglement issues crucial for quantum computing can be
understood. To analyze the differences between single-ion and
global anisotropy in FM systems,
 we found exact expressions  for the single-ion spin
matrix elements of four general spins, and  compared global and
single-ion anisotropies in the Hartree approximation for the
magnetization, specific heat, EPR and INS transitions for
equal-spin SMM tetramers with molecular group $T_d$, $D_{2d}$, and
$C_{4v}$ symmetries. Surprisingly, we also found that each
molecular group symmetry generates site-dependent single-ion
anisotropy, and that azimuthal $D_{2d}$ single-ion anisotropy has
a continuous symmetry, observed as splittings in the Ni$_4$ EPR
resonances.\cite{Edwards}

\begin{figure}
\includegraphics[width=0.17\textwidth]{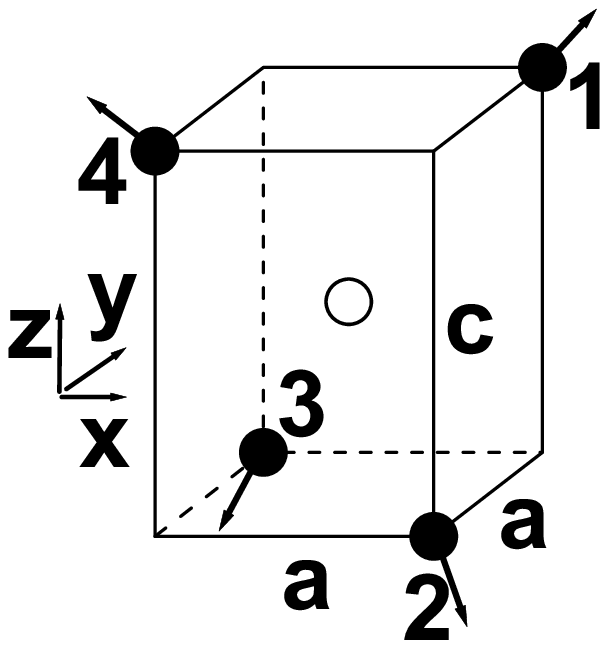}\hskip10pt\includegraphics[width=0.15\textwidth]{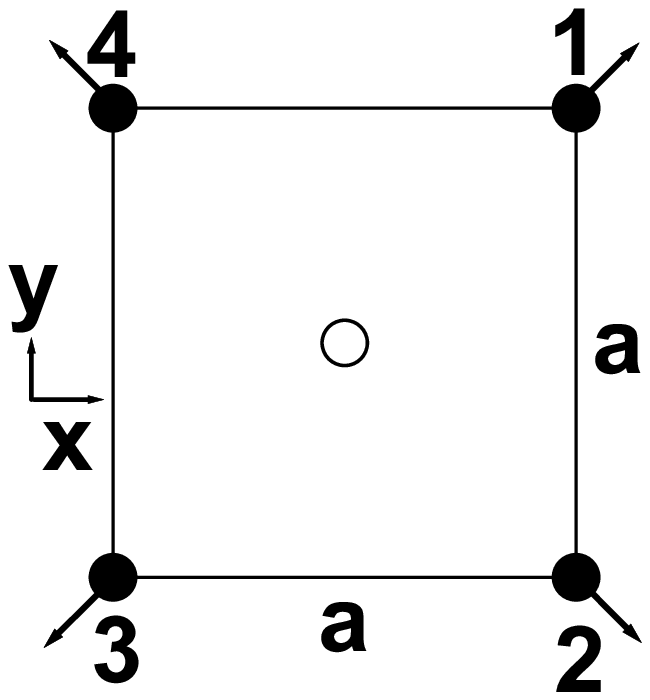}
\caption{$D_{2d}$ (left) and $C_{4v}$ (right) ion sites (filled).
Circle:  origin.  Arrows:  local axial (left), azimuthal (right)
vectors.}\label{fig1}
\end{figure}

We assume  four equal spins $s_1$ sit  on opposite corners of an
orthorhombic prism with sides $(a,a,c)$ or on a square, with the
geometric center at the origin, as in Fig. 1. For molecular
 (site point) groups $g=T_d$, $D_{2d}$, and
$C_{4v}$ with $\sum_{n=1}^4{\bm r}_n=0$, the relative spin
positions are
\begin{eqnarray}
{\bm r}_{n}&=&-\frac{a}{2}[\gamma_n^{+}\hat{\bm x}+(-1)^n\hat{\bm
y}]+\frac{c}{2}\gamma_n^{-}\hat{\bm z},\label{rn}\\
\gamma_n^{\pm}&=&\epsilon_n^{+}(-1)^{n/2}\pm\epsilon_n^{-}(-1)^{(n+1)/2},\label{gamman}
\end{eqnarray}
where $\epsilon^{\pm}_n=[1\pm(-1)^n]/2$.\cite{Tinkham} In
tetrahedrons with $g=T_d$, $c/a=1$, approximately as in  Co$_4$
and Cr$_4$.\cite{Co4,Cr4} In squares with $g=C_{4v}$, $c=0$,  as
in one Mn$_4$ SMM and Nd$_4$ (with equal total angular momentum
$j=9/2$).\cite{Boskovic,Nd4,KlemmLuban} In prisms with $g=D_{2d}$,
$c/a>1$, approximately as in Ni$_4$, a Fe$_4$ and a
Cu$_4$,\cite{Ni4} or $c/a<1$.  $\hat{\bm x}, \hat{\bm y}, \hat{\bm
z}$ are the molecular (or global) axes of each tetramer SMM.

 The local (or single-ion) azimuthal vectors satisfying all $g=C_{4v}$ symmetries
are $\hat{\bm x}^{g}_n=(-\gamma_n^{+}\hat{\bm
x}+\gamma_n^{-}\hat{\bm y})/\sqrt{2}$, as in the right panel of
Fig. 1, and the common local axial vector is $\hat{\bm
z}^{g}_n=\hat{\bm z}$. For $g=T_d, D_{2d}$, we take the group
symmetry-satisfying local axial
 vectors to be
$\hat{\bm z}^{g}_n={\bm r}_n/a_0$, as in the left panel of Fig. 1,
where $a_0=\sqrt{2a^2+c^2}$, and we set $\overline{c}=c/a_0$ and
$\overline{a}=a/a_0=\sqrt{(1-\overline{c}^2)/2}$. We define the
local azimuthal vectors from $(\hat{\bm
z}^{g}_1)^{T}=(\overline{a},\overline{a},\overline{c})$,
\begin{eqnarray}
\hat{\bm
x}^{g}_1&=&\frac{1}{2}\left(\begin{array}{c}-(1+\overline{c})\cos\mu+(1-\overline{c})\sin\mu\\
(1-\overline{c})\cos\mu-(1+\overline{c})\sin\mu\\
2\overline{a}(\sin\mu+\cos\mu)\end{array}\right),\label{x1}
\end{eqnarray}
and $\hat{\bm y}^g_1=\hat{\bm z}^g_1\times\hat{\bm x}^g_1$.  For
$\mu=0$, $\hat{\bm x}^g_1\cdot\hat{\bm z}^g_1=0$.  A rotation by
the arbitrary angle $\mu$ about $\hat{\bm z}^g_1$ leads to Eq.
(\ref{x1}).  The other local azimuthal vectors are obtained by
$\pi$ rotations of $\hat{\bm x}_1^g, \hat{\bm y}_1^g$ about
$\hat{\bm x}, \hat{\bm y}, \hat{\bm z}$. They automatically
satisfy all of the mirror planes of $D_{2d}$. \cite{Tinkham}
Although $\mu$ is a degree of freedom in $D_{2d}$, no  $\mu$
choice satisfies the remaining group operations (rotations by
$\pm2\pi/3$ about the cube diagonals) of $T_d$. Hence, $T_d$ only
has local axial vectors.

In the local coordinates of groups $g=T_d$, $D_{2d}$, $C_{4v}$,
the most general quadratic single-ion Hamiltonian is
\begin{eqnarray}
{\cal H}^g_{si}&=&-\sum_{n}\Bigl(J_a({\bm S}_n\cdot\hat{\bm
z}^g_n)^2\nonumber\\
& &+J_e[({\bm S}_n\cdot\hat{\bm x}^g_n)^2-({\bm S}_n\cdot\hat{\bm
y}^g_n)^2]\Bigr),\label{Hsi}
\end{eqnarray}
which for these equal-spin, high symmetry systems has
site-independent $J_a, J_e$.   ${\cal H}^g_{si}$ is invariant
under all allowed $g$ symmetries.  Unequal spin values or ligands
or local structural distortions from the ideal molecular group
symmetry, as commonly occur in real systems,\cite{NaV2O5} break
these $g$ symmetries, and also lead to Dzyaloshinshii-Moriya (DM)
interactions, which vanish for
 precise molecular groups $T_d$,  $D_{2d}$, or $C_{4v}$.\cite{Moriya,Waldmann}
Such lower symmetry cases will be discussed
elsewhere.\cite{ekfuture}

To make contact with experiment, we rewrite ${\cal H}^g_{si}$ in
the molecular $(\hat{\bm x},\hat{\bm y},\hat{\bm z})$
representation,
\begin{eqnarray}\tilde{\cal
H}^g_{si}&=&-\sum_{n}\Bigl(J_z^g(\mu)S_{n,z}^2+J_{xy}^g(\mu)(S_{n,x}^2-S_{n,y}^2)\nonumber\\
&
&+\sum_{\alpha\ne\beta}J_{n,\alpha\beta}^g(\mu)\{S_{n,\alpha},S_{n,\beta}\}\Bigr),\label{molecular}\\
J_z^{D_{2d}}(\mu)&=&J_a(\overline{c}^2-\overline{a}^2)+3J_e\overline{a}^2\sin(2\mu),\\
J_{xy}^{D_{2d}}(\mu)&=&J_e\overline{c}\cos(2\mu),\label{Jofmu}
\end{eqnarray} where $\alpha,\beta=x,y,z$, and the remaining non-vanishing couplings are
$J_z^{C_{4v}}=J_a-J_e/2$, $J_{n,xy}^{C_{4v}}=J_e(-1)^n/2$,
$J_{n,xy}^{D_{2d}}(\mu)=\gamma_n^{-}[J_a\overline{a}^2+J_e(1-\overline{a}^2)\sin(2\mu)]$,
$J_{n,xz}^{D_{2d}}(\mu)= J_n^{+}(\mu)$,
$J_{n,yz}^{D_{2d}}=-\gamma_n^{-}J_n^{-}(\mu)$,
$J_n^{\pm}(\mu)=(-1)^n\overline{a}\bigl(J_a\overline{c}+J_e[\cos(2\mu)\pm\overline{c}\sin(2\mu)]\bigr)$.
 In Eq. (\ref{molecular}), $\{A,B\}=AB+BA$ and we subtracted an
irrelevant constant. To preserve the $g$ symmetries of ${\cal
H}^g_{si}$, $\tilde{\cal H}^g_{si}$ contains the site-dependent
interactions $J^g_{n,\alpha\beta}(\mu)$. The continuous
(rotational) symmetry of the local azimuthal vectors in $D_{2d}$
leads to anisotropy energy strengths in $\tilde{\cal H}_{si}^g$
that are periodic functions of $\mu$, as in one-dimensional
optical phonon bands.

We assume a molecular Hamiltonian of ${\cal H}={\cal
H}_0+\tilde{\cal H}^g_{si}+{\cal H}_p$, where the Heisenberg and
Zeeman interactions are
\begin{eqnarray}
{\cal H}_0&=&-J{\bm S}^2/2-\gamma{\bm B}\cdot{\bm S}-\delta
J^g({\bm S}_{13}^2+{\bm S}_{24}^2)/2,\label{H0}
\end{eqnarray}
where $\delta J^{g}=0, J'-J,$ and $-J$ for $g=T_{d}, D_{2d}$, and
$C_{4v}$ respectively, ${\bm S}_{13}={\bm S}_1+{\bm S}_3$, ${\bm
S}_{24}={\bm S}_2+{\bm S}_4$, ${\bm S}={\bm S}_{13}+{\bm S}_{24}$,
$\gamma=g\mu_B$ is the gyromagnetic ratio (assumed isotropic, with
$g\approx2$),  and ${\bm
B}=B(\sin\theta\cos\phi,\sin\theta\sin\phi,\cos\theta)$ is the
magnetic induction at an arbitrary direction $(\theta,\phi)$
relative to the molecular coordinates. The time correlation
functions of the classical analog of this $C_{4v}$  ${\cal H}_0$
model were published.\cite{KlemmLuban}

 The phenomenological
global anisotropy interactions usually studied in SMM's  are
\begin{eqnarray}
{\cal H}_p&=&-J_bS_z^2-J_d(S_x^2-S_y^2),
\end{eqnarray}
containing axial and azimuthal contributions,
respectively.\cite{general} They are generally defined relative to
the global spin principal axes, which for equal spin, high
symmetry systems are the molecular axis vectors.

To take proper account of ${\bm B}$ in ${\cal H}_0$, we construct
our SMM eigenstates in the induction representation by
$(\hat{\tilde{\bm x}},\hat{\tilde{\bm y}},\hat{\tilde{\bm
z}})=(\hat{\bm x},\hat{\bm y},\hat{\bm z})\cdot\tensor{M}$ so that
${\bm B}=B\hat{\tilde{\bm z}}$,\cite{ek2} where
\begin{eqnarray}
\tensor{M}&=&\left(\begin{array}{ccc}\cos\theta\cos\phi &\cos\theta\sin\phi &-\sin\theta\\
-\sin\phi &\cos\phi &0\\
\sin\theta\cos\phi
&\sin\theta\sin\phi&\cos\theta\end{array}\right).\label{rotation}
\end{eqnarray}
A subsequent arbitrary rotation about $\hat{\tilde{\bm z}}$ does
not affect the eigenstates.\cite{ek2} We then set $\hbar=1$ and
write
\begin{eqnarray}
{\bm S}^2|\psi_{s,m}^{s_{13},s_{24}}\rangle&=&s(s+1)|\psi_{s,m}^{s_{13},s_{24}}\rangle,\label{S}\\
{\bm S}_{13}^2|\psi_{s,m}^{s_{13},s_{24}}\rangle&=&s_{13}(s_{13}+1)|\psi_{s,m}^{s_{13},s_{24}}\rangle,\\
{\bm S}_{24}^2|\psi_{s,m}^{s_{13},s_{24}}\rangle&=&s_{24}(s_{24}+1)|\psi_{s,m}^{s_{13},s_{24}}\rangle,\\
S_{\tilde{z}}|\psi_{s,m}^{s_{13},s_{24}}\rangle&=&m|\psi_{s,m}^{s_{13},s_{24}}\rangle,\label{Sz}\\
 S_{\tilde{\sigma}}|\psi_{s,m}^{s_{13},s_{24}}\rangle&=&A_s^{\tilde{\sigma}
m}|\psi_{s,m+\tilde{\sigma}}^{s_{13},s_{24}}\rangle,\label{SxSy}\\
A_s^{m}&=&\sqrt{(s-m)(s+m+1)},\label{Asm}
\end{eqnarray}
 where $S_{\tilde{\sigma}}=S_{\tilde{x}}+i\tilde{\sigma} S_{\tilde{y}}$ with $\tilde{\sigma}=\pm$.
For brevity, we denote $\nu=\{s,m,s_{13},s_{24},\{s_n\}\}$, and
write $|\nu\rangle\equiv|\psi_{s,m}^{s_{13},s_{24}}\rangle$. From
Eqs. (\ref{S}) to (\ref{Sz}), $\langle\nu'|{\cal
H}_0|\nu\rangle=E_{\nu,0}\delta_{\nu',\nu}$, where
\begin{eqnarray}
E_{\nu,0}&=&-Js(s+1)/2-\gamma
Bm\nonumber\\
& &-\delta J^g[s_{13}(s_{13}+1)+s_{24}(s_{24}+1)]/2.\label{E0}
\end{eqnarray}
We then transform $\tilde{\cal H}_{si}^g$ and ${\cal H}_p$ to the
induction representation, and make a standard perturbation
expansion for small anisotropy energies
$\{J_j\}=(J_a,J_b,J_d,J_e)$ relative to $|J|$ and $\gamma
B$.\cite{ek2} The  matrix elements of ${\cal H}_p$ are then
obtained from Eqs. (\ref{Sz}) and (\ref{SxSy}).   However, the
matrix elements of  $\tilde{\cal H}_{si}^g$ contain more
interesting physics.

By using symbolic manipulation software for the Clebsch-Gordan
algebra,
 we find  the  single-ion spin
matrix elements  with  general $\{s_n\}=(s_1,s_2,s_3,s_4)$ to be
\begin{eqnarray}
\langle\nu'|S_{n,\tilde{z}}|\nu\rangle
&=&\delta_{m',m}\biggl(m\delta_{s',s}\Gamma_{s^{}_{13},s_{13}',s^{}_{24},s_{24}'}^{\{s_n\},s,i}\nonumber\\
&
&+\delta_{s',s+1}C_{-s-1}^m\Delta_{s^{}_{13},s_{13}',s^{}_{24},s_{24}'}^{\{s_n\},-s-1,n}\nonumber\\
& &+\delta_{s',s-1}C_s^m\Delta_{s^{}_{13},s_{13}',s^{}_{24},s_{24}'}^{\{s_n\},s,n}\biggr),\label{Mz}\\
\langle\nu'|S_{n,\tilde{\sigma}}|\nu\rangle&=&\delta_{m',m+\tilde{\sigma}}\biggl(A_s^{\tilde{\sigma}
m}\delta_{s',s}
\Gamma_{s^{}_{13},s_{13}',s^{}_{24},s_{24}'}^{\{s_n\},s,n}\nonumber\\
&
&-\delta_{s',s+1}D_{-s-1}^{\tilde{\sigma},m}\Delta_{s^{}_{13},s_{13}',s^{}_{24},s_{24}'}^{\{s_n\},-s-1,n}\nonumber\\
& &+\delta_{s',s-1}D_s^{\tilde{\sigma},m}\Delta_{s^{}_{13},s_{13}',s^{}_{24},s_{24}'}^{\{s_n\},s,n}\biggr),\label{Msigma}\\
C_s^m&=&\sqrt{s^2-m^2},\\
D_s^{\tilde{\sigma},m}&=&\tilde{\sigma}\sqrt{(s-\tilde{\sigma}m)(s-\tilde{\sigma}m-1)},\\
\Gamma_{s_{13},s_{13}',s_{24},s_{24}'}^{\{s_n\},s,n}&=&\delta_{s_{24}',s^{}_{24}}\epsilon_n^{-}\alpha_{s_1,s_3}^{s^{}_{24},s,n}(s^{}_{13},s_{13}')\nonumber\\
&
&+\delta_{s_{13}',s^{}_{13}}\epsilon_n^{+}\alpha_{s_2,s_4}^{s_{13},s,n}(s_{24},s_{24}'),\\
\Delta_{s^{}_{13},s_{13}',s^{}_{24},s_{24}'}^{\{s_n\},s,n}&=&\delta_{s_{24}',s^{}_{24}}\epsilon_n^{-}\beta_{s_1,s_3}^{s_{24},s,n}(s^{}_{13},s_{13}')\nonumber\\
&
&+\delta_{s_{13}',s^{}_{13}}\epsilon_n^{+}\beta_{s_2,s_4}^{s^{}_{13},s,n}(s^{}_{24},s_{24}'),\\
\alpha_{s_1,s_3}^{s^{}_{24},s,n}(s^{}_{13},s_{13}')&=&\frac{1}{4}(1+\xi_{s,s^{}_{13},s^{}_{24}})\delta_{s_{13}',s^{}_{13}}\nonumber\\
&
&-\gamma_n^{+}\Bigl(F^{s_{13},s_{24}}_{s_1,s_3,s}\delta_{s_{13}',s^{}_{13}-1}\nonumber\\
& &+F^{s_{13}+1,s_{24}}_{s_1,s_3,s}\delta_{s_{13}',s_{13}+1}\Bigr),\\
\beta_{s_1,s_3}^{s^{}_{24},s,n}(s_{13},s_{13}')&=&-\frac{(-1)^n}{4}\eta_{s,s_{13},s_{24}}\delta_{s_{13}',s^{}_{13}}\nonumber\\
&
&-\gamma_n^{+}\Bigl(G_{s_1,s_3,s}^{s_{13},s_{24}}\delta_{s_{13}',s^{}_{13}-1}\nonumber\\
&
&+G_{s_1,s_3,-s}^{s_{13}+1,s_{24}}\delta_{s_{13}',s^{}_{13}+1}\Bigr),\\
F_{s_1,s_3,s}^{s_{13},s_{24}}&=&-\frac{\eta_{s_{13},s_1,s_3}A_{s+s_{13}}^{s_{24}}A_{s_{24}}^{s-s_{13}}}{4s(s+1)},\label{A}\nonumber\\
& &\\
G_{s_1,s_3,s}^{s_{13},s_{24}}&=&\frac{\eta_{s_{13},s_1,s_3}A_{s+s_{13}}^{s_{24}}A_{s+s_{13}-1}^{s_{24}}}{4s\sqrt{4s^2-1}},\label{B}\nonumber\\
& &\\
\eta_{z,x,y}&=&\frac{A_{x+z}^yA_y^{x-z}}{\sqrt{z^2(4z^2-1)}},\label{eta}\nonumber\\
& &\\
 \xi_{z,x,y}&=&\frac{x(x+1)-y(y+1)}{z(z+1)},\label{xi}
\end{eqnarray}
where $\gamma_n^{+}$ is given by Eq. (\ref{gamman}). The
prefactors $m$, $A_s^{\tilde{\sigma} m}$, $C_s^m$, $C_{-s-1}^m$,
$D_s^{\tilde{\sigma},m}$, and $D_{-s-1}^{\tilde{\sigma},m}$ are
consequences of the Wigner-Eckart theorem for a vector
operator.\cite{Tinkham} The challenge was to obtain the
coefficients
$\Gamma_{s^{}_{13},s_{13}',s^{}_{24},s_{24}'}^{\{s_n\},s,n}$ and
$\Delta_{s^{}_{13},s_{13}',s^{}_{24},s_{24}'}^{\{s_n\},s,n}$.
Their hierarchical structure based upon  the unequal-spin dimer
suggests that analogous coefficients  with $n
> 4$ may be obtainable.\cite{ek2}

 At arbitrary $(\theta,\phi)$, the first order
corrections $E_{\nu,1}^{\mu}=\langle\nu|{\cal H}_{p}+\tilde{\cal
H}^g_{si}|\nu\rangle$ to the eigenstate energies are
\begin{eqnarray}
E_{\nu,1}&=&\frac{\tilde{J}^{g,\overline{\nu}}_{b,z}(\mu)}{2}[m^2-s(s+1)]-J_z^g(\mu)b_{\overline{\nu}}\nonumber\\
&
&-\frac{[3m^2-s(s+1)]}{2}\Bigl(\tilde{J}^{g,\overline{\nu}}_{b,z}(\mu)\cos^2\theta\nonumber\\
& &+\tilde{J}^{g,\overline{\nu}}_{d,xy}(\mu)\sin^2\theta\cos(2\phi)\Bigr),\label{E1}\\
\tilde{J}_{b,z}^{g,\overline{\nu}}(\mu)&=&J_b+J_z^g(\mu)a_{\overline{\nu}},\\
\tilde{J}_{d,xy}^{g,\overline{\nu}}(\mu)&=&J_d+{J}^g_{xy}(\mu)a_{\overline{\nu}},\\
a_{\overline{\nu}}&=&\frac{1}{4}\Bigl(1+\xi^2_{s,s_{13},s_{24}}-\eta_{s,s_{13},s_{24}}^2-\eta_{s+1,s_{13},s_{24}}^2\Bigr)\nonumber\\
&
&+2\biggl(\sum_{\sigma=\pm1}\Bigl[\Bigl(F_{s_1,s_1,s}^{s_{13}+(\sigma+1)/2,s_{24}}\Bigr)^2\nonumber\\
&
&-\sum_{\sigma'=\pm1}\Bigl(G_{s_1,s_1,\sigma\sigma's+\sigma(1+\sigma')/2}^{s_{13}+(1+\sigma)/2,s_{24}}\Bigr)^2\Bigr]\nonumber\\
&
&+(s_{13}\leftrightarrow s_{24})\biggr),\\
b_{\overline{\nu}}&=&\frac{1}{8}\sum_{\sigma'=\pm1}(2s+1+\sigma')^2\biggl(\eta^2_{s+(1+\sigma')/2,s_{13},s_{24}}\nonumber\\
&
&+8\sum_{\sigma=\pm1}\Bigl[\Bigl(G_{s_1,s_1,\sigma\sigma's+\sigma(1+\sigma')/2}^{s_{13}+(1+\sigma)/2,s_{24}}\Bigr)^2\label{b}\nonumber\\
& &+(s_{13}\leftrightarrow s_{24})\Bigr]\biggr),
\end{eqnarray}
where the interactions are given by and following Eq.
(\ref{Jofmu}), and $\overline{\nu}=\{s,s_{13},s_{24},s_1\}$
excludes $m$.
   Second order corrections
to the energies, including the  symmetry-induced site-dependent
interactions in Eq. (\ref{molecular}), will be presented
elsewhere.\cite{ekfuture} For $T_d$, these are  important, because
their site-independent single-ion interactions vanish.

For $T_d$, $D_{2d}$, or $C_{4v}$ symmetries, $E_{\nu,1}$ has a
form entirely analogous to that of the dimer,\cite{ek2} except for
the periodic $\mu$-dependence of the $D_{2d}$ interactions, which
broadens the discrete energy levels into bands, akin to
one-dimensional optical phonon modes.   This and  the strongly
different dependencies upon the quantum numbers $\overline{\nu}$
of the coefficients of the single-ion and global anisotropy
interactions can be employed to provide a definitive measure of
$J$, $\delta J^g$, and $\{J_j\}$. This allows us to determine how
well the phenomenological global anisotropy parameters simulate
the microscopic systems.  We therefore present the Hartree
approximation in the induction representation of four measurable
quantities.

The Hartree approximation is accurate at low $k_BT/|J|$ and
$\gamma B/|J|$ not too small,\cite{ek2} where $k_B$ is Boltzmann's
constant.  In this approximation, $E_{\nu}=E_{\nu,0}+E_{\nu,1}$ is
given by Eqs. (\ref{E0}) and (\ref{E1}), respectively. We define
${\rm Tr}^g_{\nu}\equiv\sum_{\nu}\int_{E_{\nu,\rm
min}}^{E_{\nu,\rm max}}{\cal D}^g_{\nu}(\epsilon)d\epsilon$.  For
$g=D_{2d}$, we set
$E_{\nu}=E_{\nu+}+E_{\nu-}\sin(2\mu+\chi_{\nu})$, where
$E_{\nu\pm}=(E_{\nu,{\rm max}}\pm E_{\nu,{\rm min}})/2$. Then
${\cal D}^g_{\nu}(\epsilon)=\frac{1}{2}[(\epsilon-E_{\nu,{\rm
min}})(E_{\nu,{\rm max}}-\epsilon)]^{-1/2}$ is sharply peaked at
$E_{\nu,{\rm max}}, E_{\nu,{\rm min}}$.  For $g=T_d,C_{4v}$,
${\cal D}^g_{\nu}(\epsilon)=\delta(\epsilon-E_{\nu})$ and
$E_{\nu,{\rm max}}>E_{\nu}>E_{\nu,{\rm min}}$.  The partition
function in the Hartree approximation is then $Z_{g}^{(1)}={\rm
Tr}^g_{\nu}e^{-\beta \epsilon}$, where $\beta=1/(k_BT)$,
 $|s_1-s_3|\le s_{13}\le
s_1+s_3$, $|s_2-s_4|\le s_{24}\le s_2+s_4$, $|s_{13}-s_{24}|\le
s\le s_{13}+s_{24}$, and $-s\le m\le s$.  Then,  the Hartree
magnetization $M^{(1)}_g({\bm B},T)=\gamma{\rm
Tr}^g_{\nu}\bigl(me^{-\beta\epsilon}\bigr)/Z_g^{(1)}$ and specific
heat
\begin{eqnarray}
\frac{C_{g,V}^{(1)}}{k_B\beta^2}&=&\frac{{\rm Tr}^g_{\nu}(
\epsilon^2e^{-\beta\epsilon})}{Z_g^{(1)}}-\Bigl(\frac{{\rm
Tr}_{\nu}^g(\epsilon e^{-\beta\epsilon})}{Z_g^{(1)}}\Bigr)^2.
\end{eqnarray}

The Hartree EPR absorption
${\Im}\chi_{-\sigma,\sigma}^{g,(1)}({\bm B},\omega)$ for
clockwise ($\sigma=1$) or counterclockwise ($\sigma=-1$)
circularly polarized oscillatory fields normal to ${\bm B}$  is
\begin{eqnarray}
{\Im}\chi^{g,(1)}_{-\sigma,\sigma}&=&\frac{\gamma^2}{Z_g^{(1)}}{\rm
Tr}^g_{\nu}{\rm Tr}^g_{\nu'} e^{-\beta\epsilon}\bigl|M_{\nu,\nu'}\bigr|^2\nonumber\\
&
&\times\bigl[\delta(\epsilon-\epsilon'+\omega)-\delta(\epsilon'-\epsilon+\omega)\bigr],
\end{eqnarray}
where  $M_{\nu,\nu'}=A_s^{\sigma
m}\delta_{m',m+\sigma}\delta_{s',s}\delta_{s_{13}',s^{}_{13}}\delta_{s_{24}',s^{}_{24}}$
and ${\rm Tr}^g_{\nu'}=\sum_{\nu'}\int_{E_{\nu',\rm
min}}^{E_{\nu',\rm max}}{\cal D}^g_{\nu'}(\epsilon')d\epsilon'$.
The resonant inductions  for $g=T_d,C_{4v}$ are strong and
centered at
\begin{eqnarray}
\gamma B^{(1)}_{\rm
res}&=&\pm\omega+\frac{(2m+\sigma)}{2}\Bigl((1-3\cos^2\theta)\tilde{J}_{b,z}^{g,\overline{\nu}}(0)\nonumber\\
&
&-3\sin^2\theta\cos(2\phi)\tilde{J}_{d,xy}^{g,\overline{\nu}}(\pi/4)\Bigr).\label{Bres}
\end{eqnarray}
 For $g=D_{2d}$, the weak center of each $B^{(1)}_{\rm res}$,
 Eq. (\ref{Bres}),
 is equally
surrounded by two strong resonances  split by $\gamma\Delta
B^{(1)}_{\rm res}=2|\sigma J_e(z_{\nu'}-z_{\nu})|$, where
$z_{\nu}=\sqrt{x_{\nu}^2+y_{\nu}^2},
y_{\nu}=\frac{1}{2}\overline{c}a_{\overline{\nu}}[3m^2-s(s+1)]\sin^2\theta\cos(2\phi)$,
and $x_{\nu}=\frac{3}{2}\overline{a}^2\{2b_{\overline{\nu}}+
a_{\overline{\nu}}[s(s+1)\sin^2\theta+m^2(3\cos^2\theta-1)]\}.$
 For $\theta=0$, $\gamma\Delta B^{(1)}_{\rm
res}=6\overline{a}^2|(2m+\sigma)a_{\overline{\nu}}J_e|$, nicely
fitting $s=4$ $(a_{\overline{\nu}}=1/7)$ data on three Ni$_4$
compounds.  For  $c/a\approx1.1$, $|J_e|\approx$ 0.07K, 0.63K, and
0.91K.\cite{Edwards} More $B_{\rm res}(\theta,\phi)$ data are
urged.

The Hartree INS cross-section $S_g^{(1)}({\bm B},{\bm q},\omega)$
is
\begin{eqnarray}
S_g^{(1)}&=&{\rm Tr}^g_{\nu}\sum_{\nu'}e^{-\beta\epsilon}\sum_{\tilde{\alpha},\tilde{\beta}}\bigl(\delta_{\tilde{\alpha},\tilde{\beta}}-\hat{q}_{\tilde{\alpha}}\hat{q}_{\tilde{\beta}}\bigr)\sum_{n,n'=1}^4\nonumber\\
& &\times e^{i{\bm q}\cdot({\bm r}_n-{\bm
r}_{n'})}\langle\nu|S_{n',\tilde{\alpha}}^{\dag}|\nu'\rangle\langle\nu'|S_{n,\tilde{\beta}}|\nu\rangle,
\end{eqnarray}
where
$\tilde{\alpha},\tilde{\beta}=\tilde{x},\tilde{y},\tilde{z}$,
$\hat{q}_{\tilde{x}}=\sin\theta_{b,q}\cos\phi_{b,q}$,
$\hat{q}_{\tilde{y}}=\sin\theta_{b,q}\sin\phi_{b,q}$, and
$\hat{q}_{\tilde{z}}=\cos\theta_{b,q}$,  $\theta_{b,q}$ and
$\phi_{b,q}$ describe the  relative orientations of ${\bm B}$ and
${\bm q}$,\cite{ek2}  the ${\bm r}_n$ and
$\langle\nu'|S_{n,\tilde{\alpha}}|\nu\rangle$ are given by Eqs.
(\ref{rn}), (\ref{Mz}), and (\ref{Msigma}) respectively, and the
scalar ${\bm q}\cdot({\bm r}_n-{\bm r}_{n'})$ is invariant under
the rotation, Eq. (\ref{rotation}). As for the dimer,\cite{ek2}
additional EPR and INS transitions with amplitudes higher order in
the $\{J_j\}$ are obtained in the extended Hartree approximation,
but will be presented elsewhere for brevity.\cite{ekfuture}

We presented a microscopic theory of high-symmetry single molecule
magnets. From our exact single-ion spin matrix elements for four
general spins, we studied the most general quadratic single-ion
interactions in equal-spin tetramers with molecular group symmetry
$T_d$, $D_{2d}$, or $C_{4v}$. Each group introduces site-dependent
single-ion interactions, and azimuthal anisotropy with $D_{2d}$
symmetry has a continuous symmetry observed as splittings in
Ni$_4$ EPR resonances, providing a direct measure of the
microscopic interaction. We used the Hartree approximation to
provide explicit expressions for the magnetization, specific heat,
EPR absorption, and INS cross-section,  valid at low temperatures
and sufficiently large magnetic fields.  Our procedure is
extendable to  systems with lower symmetry, higher-order
single-ion interactions, and possibly to systems with $n>4$ spins.

This work was supported by the NSF under contract NER-0304665.

\end{document}